\documentclass[aip,jcp,amsmath,amssymb,preprint]{revtex4-1}
\usepackage{graphicx}
\usepackage{dcolumn}
\usepackage[version=3]{mhchem}

\begin{document}

\title{Penning Spectroscopy and Structure of Acetylene Oligomers in \ce{He} Nanodroplets}
\author{S. Mandal}
\affiliation {Indian Institute of Science Education and Research, Pune~411008, India}

\author{R. Gopal}
\affiliation {Tata Institute of Fundamental Research, Hyderabad~500107, India}

\author{M. Shcherbinin}
 \affiliation {Aarhus University, 8000 Aarhus C, Denmark}
 
\author{A. D’Elia}
\affiliation {Department of Physics, University of Trieste, Via A. Valerio 2, 34127 Trieste, Italy}

\author{H. Srinivas}
\affiliation {Max-Planck-Institute f\"{u}r Kernphysik, 69117 Heidelberg, Germany}

\author{R. Richter}
\affiliation {Elettra-Sincrotrone Trieste, 34149 Basovizza, Italy}

\author{M. Coreno}
\affiliation {Elettra-Sincrotrone Trieste, 34149 Basovizza, Italy}
\affiliation {Consiglio Nazionale delle Ricerche – Istituto di Struttura della Materia, 34149 Trieste, Italy}

\author{B. Bapat}
\affiliation {Indian Institute of Science Education and Research, Pune~411008, India}

\author{M. Mudrich}
\affiliation {Aarhus University, 8000 Aarhus C, Denmark}
\affiliation {Indian Institute of Technology Madras, Chennai 600036, India}

\author{S. R. Krishnan}
\email[]{srkrishnan@iitm.ac.in}
\affiliation {Indian Institute of Technology Madras, Chennai 600036, India}

\author{V. Sharma}
\email[]{vsharma@phy.iith.ac.in}
 \affiliation {Indian Institute of Technology Hyderabad, Kandi~502285, India}

\date{13-April-2020}

\begin{abstract}
Embedded atoms or molecules in a photoexcited \ce{He} nanodroplet are well-known to be ionized through inter-atomic relaxation in a  Penning process. In this work, we investigate the Penning ionization of acetylene oligomers occurring from the  photoexcitation bands of \ce{He} nanodroplets.
In close analogy to conventional Penning electron spectroscopy by thermal atomic collisions, the  $n=2$ photoexcitation band plays the role of the metastable atomic $1s2s$ $^{3,1}S$ \ce{He}$^\ast$. This facilitates electron spectroscopy of acetylene aggregates in the sub-kelvin \ce{He} environment, providing the following insight into their structure: The molecules in the dopant cluster are loosely bound van der Waals complexes rather than forming covalent compounds. In addition, this work reveals a Penning process stemming from the $n=4$ band where charge-transfer from autoionized \ce{He} in the droplets is known to be the dominant relaxation channel. This allows for excited states of the remnant dopant oligomer Penning-ions to be studied.
Hence, we demonstrate Penning ionization electron spectroscopy of doped droplets as an effective technique for investigating dopant oligomers which are easily formed by attachment to the host cluster.

\end{abstract}

\maketitle

\section{Introduction}
\ce{He} nanodroplets have been regarded as an ideal host environment for spectroscopic studies of embedded atoms and molecules over a vast spectral range spanning from  the infrared to the vacuum ultraviolet due to their ability to ro-vibronically cool these dopants without any chemical modification. However, this seemingly passive \ce{He} host environment proves to be fertile ground for studying a rich class of intermolecular relaxation processes between the excited host and the attached dopants when being photoexcited \cite{buc13, laf16, laf19}. A potent observable for obtaining insights into these processes is the energy distribution of ejected electrons tagged to particular ions arising out of these multi-atomic processes. This observable can be applied specifically to study the indirect Penning ionization of dopant aggregates interacting with photoexcited \ce{He}$^\ast$ in the droplets whereby the participating quantum states of both the embedded species and the droplets can be discerned. Although recent report \cite{shc18} suggested that the scattering of electrons following  Penning ionization may obscure the molecular features, we were able to resolve the Penning ionization electron spectra (PIES) in the case of acetylene (\ce{C2H2}) oligomers.
This motivates the development of ion-correlated energy-resolved electron detection in combination with the Penning process as a spectroscopic tool to study the electronic structure of weakly-bound quantum aggregates. These atomic and molecular complexes can be aggregated with relative ease by employing He nanodroplets as a nanoscale sub-Kelvin container \cite{sch04, prz08, las19}. The application of Penning spectroscopy aided by coincident electron-ion detection to small acetylene clusters in nanodroplets indicates a loosely bound van der Waals aggregate of \ce{C2H2} molecules in the sub-Kelvin \ce{He} nanodroplet environment.

Our investigation includes a series of electron-ion coincidence measurements detailed in the next section. Aided with information about the prominent fragmentation products through $20...26$ \textrm{eV} photon energy, the kinetic energy distributions of electrons in coincidence with these ions are measured. To initiate the Penning ionization we chose $21.6$ \textrm{eV} photoexcitation corresponding to the most important $n=2$ droplet band. Typically, doped alkali atoms which reside on the droplet surface are known to be preferentially Penning ionized by this excitation of the complex \cite{luk11, buc13}. Here, we measured electron kinetic energy spectra in coincidence with the most abundant dopant cluster ions, \ce{C2H2^+}, \ce{[C2H2]_2^+} and \ce{[C2H2]_3^+}. In contrast to studies hitherto \cite{buc13}, we were also able to use the autoionizing $n=4$ droplet band for PIES of acetylene doped \ce{He} nanodroplets which can reveal acetylene cluster ions left in excited states higher than those possible in the case of the $n=2$ excitation. Not only does PIES reveal details about the dopant oligomers, the converse, the relaxation behavior of excited \ce{He}$^\ast$ in the droplet containing the dopant is also a subject of current studies. The \ce{He} droplet is expected to internally relax from the dipole allowed $1s2p$ $^{1}P$ to $1s2s$ $^{3,1}S$ \ce{He}$^\ast$ states before the Penning ionization ensues \cite{wan08, lta19, mud20}.

This work on the Penning ionization of acetylene oligomers mediated by different photo-excitation bands of \ce{He} nanodroplets is revealing in many respects:

First, we gain insights in the relaxation dynamics of excited \ce{He} nanodroplets and in the electronic states of acetylene oligomers involved in the Penning ionization process. In addition to the $n=2$ states of \ce{He}$^\ast$, higher states of \ce{He}$^\ast$ are found to induce Penning ionization of acetylene thereby accessing higher-lying states of the acetylene product ion.

Second, PIES reveals a dominant monomer-like feature even for Penning electrons tagged to the acetylene dimer and trimer ions pointing to a weakly bound van der Waals system of the aggregate inside the droplet. This is reminiscent of the foam-like structure evidenced in the case of \ce{Mg} doped into \ce{He} nanodroplets \cite{prz08}.

\section{Experimental Details}

\begin{figure}[]
	\centering
	\includegraphics[width=\textwidth]{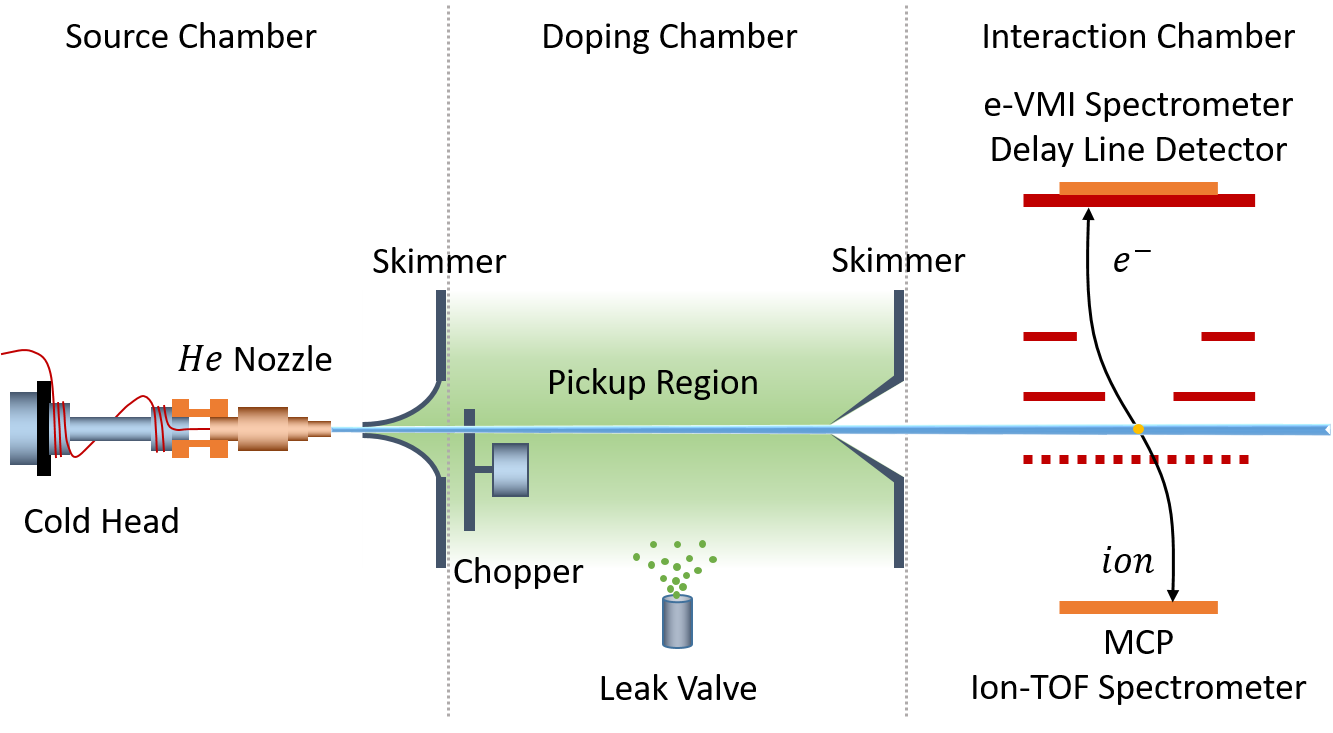}
	\caption{Schematic diagram of the experimental setup. \ce{He} droplets are generated in the source chamber by supersonic expansion of the He gas through the cryogenic nozzle and extracted into the next region by a skimmer. The jet of droplets is doped by picking up \ce{C2H2} molecules which are effused in the doping chamber downstream. Subsequently, in the interaction chamber, the doped \ce{He} droplet jet is ionized by EUV synchrotron radiation. The resultant electrons and ions are measured in coincidence by the VMI and TOF spectrometers operating in tandem.}
	\label{Fig1_Ac}
\end{figure}

Our investigations are the result of a beamtime at the Gas-Phase (GAPH) beamline at the Elettra Synchrotron Trieste, Italy. The schematic of the experiment is depicted in fig.\ref{Fig1_Ac} while the details have been presented elsewhere \cite{oke11, shc17, shc18, buc13}. In brief, the implementation consists of three sections. The first one is the source chamber where He nanodroplets are generated by supersonic expansion from a cryogenically cooled nozzle. This is attached to a doping chamber where the droplet jet picks up dopant molecules from the doping chamber. The stream of doped droplets then enters the interaction chamber where the doped droplets are intercepted by the synchrotron radiation. To produce \ce{He} nanodroplets, in the source chamber pressurized ($\sim50$ \textrm{bar}) high-purity helium gas (\ce{He} $6.0$) is supersonically expanded through a cryogenically cooled nozzle with a $5.0$ \textrm{$\mu$m} orifice. The jet is extracted using a trumpet-shaped skimmer with a $0.4$ \textrm{mm} aperture. The variation of the  nozzle temperature ($T_{noz}$) serves to control the mean size of the droplets; typically varying $T_{noz}$ between $16$ and $14$ \textrm{K} allows a control of droplet sizes between $8800$ and $23000$ \ce{He} atoms per droplet on the average \cite{fra06, toe04}.

The skimmed jet of \ce{He} nanodroplets exits the  source chamber to pick up \ce{C2H2} molecules which were effused into the doping chamber by a controlled leak through a dosing valve. This variation of the partial acetylene pressure ($P_{d}$) in this region, $6\times10^{-7}$ \textrm{mbar} to $4\times10^{-6}$ \textrm{mbar}, offers a direct control over the pick up of the dopant molecules which follows Poissonian statistics. The number of dopant molecules per droplet can be varied between at the most one dopant molecule per droplet, to several molecules captured into a typical droplet in the jet. Before doping, it is important to distil acetylene gas to remove inevitable acetone contamination. We passed the precursor through a coiled copper tube immersed in a bath with ethanol and liquid \ce{N2} slurry maintained at $173$ \textrm{K}. Furthermore, a mechanical chopper operating between the source  and the doping chambers to periodically intercept the nanodroplet jet enables us to record distinct background signals arising out of the effusive residual gas molecules in addition to acquiring signals from  doped droplets. These measurements performed in quick succession, typically switching at $\sim70$ \textrm{Hz}, allow us to reliably subtract the background due to effusive gases enabling low-noise acquisition of droplet specific signals.

Downstream of the doping chamber, the doped droplet jet  passes through a second skimmer to enter into the interaction chamber (cf. fig.\ref{Fig1_Ac}). This chamber, maintained at $\sim10^{-8}$ \textrm{mbar}  houses a velocity map imaging (VMI) spectrometer for electrons along with a time-of-flight (TOF) spectrometer for ions operating in tandem. At the geometric center of the interaction chamber the doped droplets interact with the focused beam of linearly polarized EUV photons from the synchrotron. The photon beam has a typical peak intensity of $\sim15$ \textrm{Wm$^{-2}$} at a repetition rate of $500$ \textrm{MHz} in the form of $\sim150$ \textrm{ps} pulses. We have used photon energies between $20$ and $26$ \textrm{eV} for electronic excitation and ionization of the host \ce{He} matrix. Two slits in the photon beam path were adjusted to maintain moderate count-rates in the range of $10-20$ \textrm{kHz}  on the charged particle detectors. To suppress the higher order harmonics of the synchrotron radiation, a \ce{Sn} filter was used for measurements at $21.6$ \textrm{eV} photon energy. The beamline is capable of a resolving power of $\Delta E / E \leq 10^{-4}$ over the whole photon energy range.
	
The VMI and TOF spectrometers operating  synchronously in the interaction chamber enable electron-ion coincidence measurements. Both the single and double ion coincidences with electrons were implemented in these experiments. From sufficiently long acquisitions, time-of-flight  ion-ion correlation maps were obtained along with the corresponding electron-VMI distributions. However, these ion-ion coincidence maps did not evidence any double ionization of \ce{C2H2} doped \ce{He} nanodroplets in the studied photon energy range. The kinetic energy distributions of the electrons are derived from the velocity-map-images recorded on a position sensitive delay-line detector of the VMI spectrometer. These were Abel inverted using well-established protocols - we employed  B. Dick's MEVELER \cite{dic14} for inversion. To calibrate the kinetic energy of electrons for a given configuration of the VMI spectrometer, we referenced photoionization of atomic \ce{He} over a few photon energies in the range of  $25$ \textrm{eV} to $40$ \textrm{eV}. The average energy resolution ($\Delta E/E$) of the VMI spectrometer is typically  $\sim7\%$. The resolution for the electron energy spectra shown here is limited by the resolution of the VMI spectrometer. The TOF spectra of ions were correlated to electron energy spectra obtained from the VMI.

\section{Results and Discussion}

\subsection{Ion yield}
	
\begin{figure}[]
    \centering
    \includegraphics[width=\textwidth]{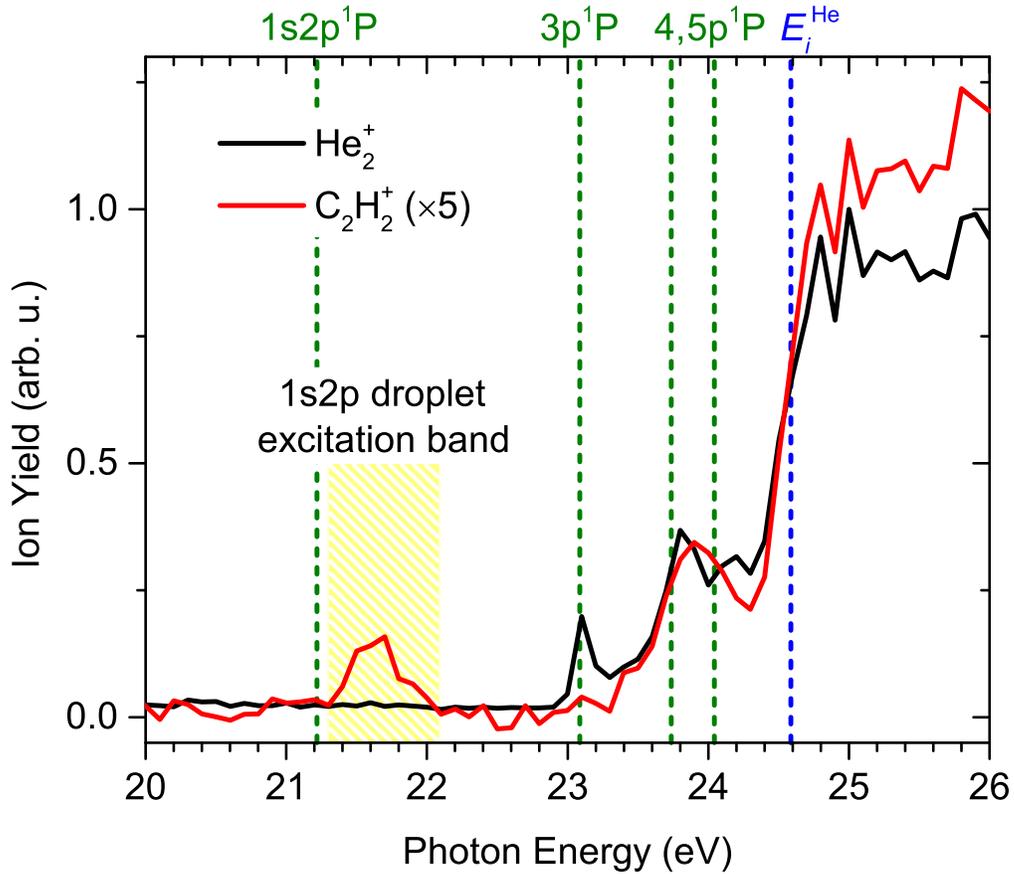}
    \caption{Ion yields of \ce{He2^+}  and \ce{C2H2^+} from the acetylene doped \ce{He} droplet as function of photon energy. The nozzle temperature and doping chamber pressure are maintained at $16$ \textrm{K} and $6.5\times10^{-7}$ \textrm{mbar}, respectively. The blue vertical dashed line represents the ionization energy of atomic \ce{He} ($E_{i}^{\ce{He}}$) and the vertical green dashed lines at $21.21$ \textrm{eV}, $23.09$ \textrm{eV}, $23.78$ \textrm{eV}, and $24.04$ \textrm{eV} represent atomic \ce{He}$^\ast$ $1s2p$ $^{1}P$, $1s3p$ $^{1}P$, $1s4p$ $^{1}P$ and $1s5p$ $^{1}P$ energy levels, respectively. The yellow shaded region shows the droplet photoexcitation band correlated to the $1s2p$ atomic \ce{He} level.}
    \label{Fig2_Ac}
\end{figure}

Droplet induced ionization of doped acetylene can be readily observed by recording the ion yield of \ce{C2H2^+} as a function of photon energy in the energy range from $20$ \textrm{eV} to $26$ \textrm{eV} (cf. fig.\ref{Fig2_Ac}). Fig.\ref{Fig2_Ac} also shows the measured \ce{He2^+} yield from doped droplets in the same photon energy range for comparison as this is known to be the most prominent ion arising out of the host. This reveals two important aspects of the induced ionization of acetylene doped droplets. The feature at the $n=2$ droplet band centered at $21.6$ \textrm{eV} where only dopant \ce{C2H2^+} ions (red) are seen without any significant yield of \ce{He2^+} is characteristic of the Penning process. At higher photon energies,  beyond the autoionization threshold ($\sim23$ \textrm{eV}) of pure \ce{He} droplets, both \ce{C2H2^+} and \ce{He2^+} yields follow  very similar trends.

It is well known that, at a photon energy of $21.6$ \textrm{eV}, there is a very high cross section for the droplet to photoexcite to the $n=2$ band derived from the $1s2p$ atomic \ce{He} level \cite{jop93}. Following this photoexcitation, due to repulsive interaction with the droplet environment, the excited \ce{He}$^\ast$ usually migrates to the surface of the droplet and Penning ionization is found to be particularly efficient for surface-bound alkali atoms \cite{buc13}. However, more recently, Penning ionization of immersed molecules was also clearly observed \cite{shc18}. As we observe a peak in the \ce{C2H2^+} yield from the droplet at the same photon energy, we expect the doped acetylene molecules, which are believed to stay at the interior of the droplet, to be ionized by the following Penning process:
	
\begin{equation}\label{Penning_ionization}
\begin{split}
	\ce{He}_{\mathit{m}} + [\ce{C2H2}]_{\mathit{n}} + h\nu & \longrightarrow \ce{He}_{\mathit{m}}^{\ast} + [\ce{C2H2}]_{\mathit{n}} \\
	& \longrightarrow \ce{He}_{\mathit{m}} + [\ce{C2H2}]_{\mathit{n}}^{+} + e^{-}_{\textrm{Penning}}
\end{split}
\end{equation}
	
The peak structures below the ionization energy of atomic \ce{He} ($E_{i}^{\ce{He}}=24.58$ \textrm{eV}) in the \ce{He2^+} yield correspond to the autoionization of the \ce{He} nanodroplets. This occurs via droplet photoexcitation to Rydberg states of \ce{He2}$^\ast$ which are derived from atomic $1s$ $np$ $^{1}P$, $n>2$ states ($23$ \textrm{eV} $<h\nu<E_{i}^{\ce{He}}$). For $h\nu \geq E_{i}^{\ce{He}}$, direct ionization of \ce{He} atom in the nanodroplets occurs. Common to both these ionization regimes, the \ce{He+} or \ce{He2^+} ion formed in the droplet usually migrates to its interior due to the net attractive interaction with rest of the droplet enabled by fast charge hopping \cite{hal98}. In the case of rare gas dopants, which reside in the droplet interior, ionization by charge-transfer is the  dominant dopant ionization process \cite{buc13}. Thus, both these regimes, autoionization and direct ionization, are expected to contribute to dopant ionization. This is convincingly evidenced by the yield of  \ce{C2H2^+} ions following the trend of the host \ce{He2^+} ion for $h\nu > 23$ \textrm{eV}, cf. fig.\ref{Fig2_Ac}, due to a charge-transfer process:
	
\begin{equation}\label{CT_ionization}
\begin{split}
	\ce{He}_{\mathit{m}} + [\ce{C2H2}]_{\mathit{n}} + h\nu & \longrightarrow \ce{He}_{\mathit{m}}^{+} + [\ce{C2H2}]_{\mathit{n}} + e^{-}_{\ce{He}} \\
	& \longrightarrow \ce{He}_{\mathit{m}} + [\ce{C2H2}]_{\mathit{n}}^{+} + e^{-}_{\ce{He}}.
\end{split}
\end{equation}
	
However, in the autoionization regime ($23$ \textrm{eV} $<h\nu<E_{i}^{\ce{He}}$), previous studies with \ce{Li} and \ce{Ar} doped \ce{He} nanodroplets reported significantly low contributions of Penning ionization from \ce{He}$^\ast$ ($1s2s$) excited states which competes with the dominant charge-transfer channel \cite{buc13}. Further, in comparison to the \ce{He}$^\ast$ $1s2s$ states, the contribution to dopant Penning ionization from higher excited states \ce{He}$^\ast$ ($1s$ $np$, $n>2$) in \ce{Li} doped droplets is rather small \cite{lta19}. The energy distributions of the emitted electrons measured in coincidence with ions for both the  Penning \eqref{Penning_ionization} and charge-transfer \eqref{CT_ionization} ionization processes are expected to be distinct. The corresponding electron spectra would also enable the identifications of the ionization processes and of the participating electronic states corresponding to the doped \ce{C2H2} oligomer and the \ce{He} host.
	
\begin{figure}[]
	\centering
	\includegraphics[width=\textwidth]{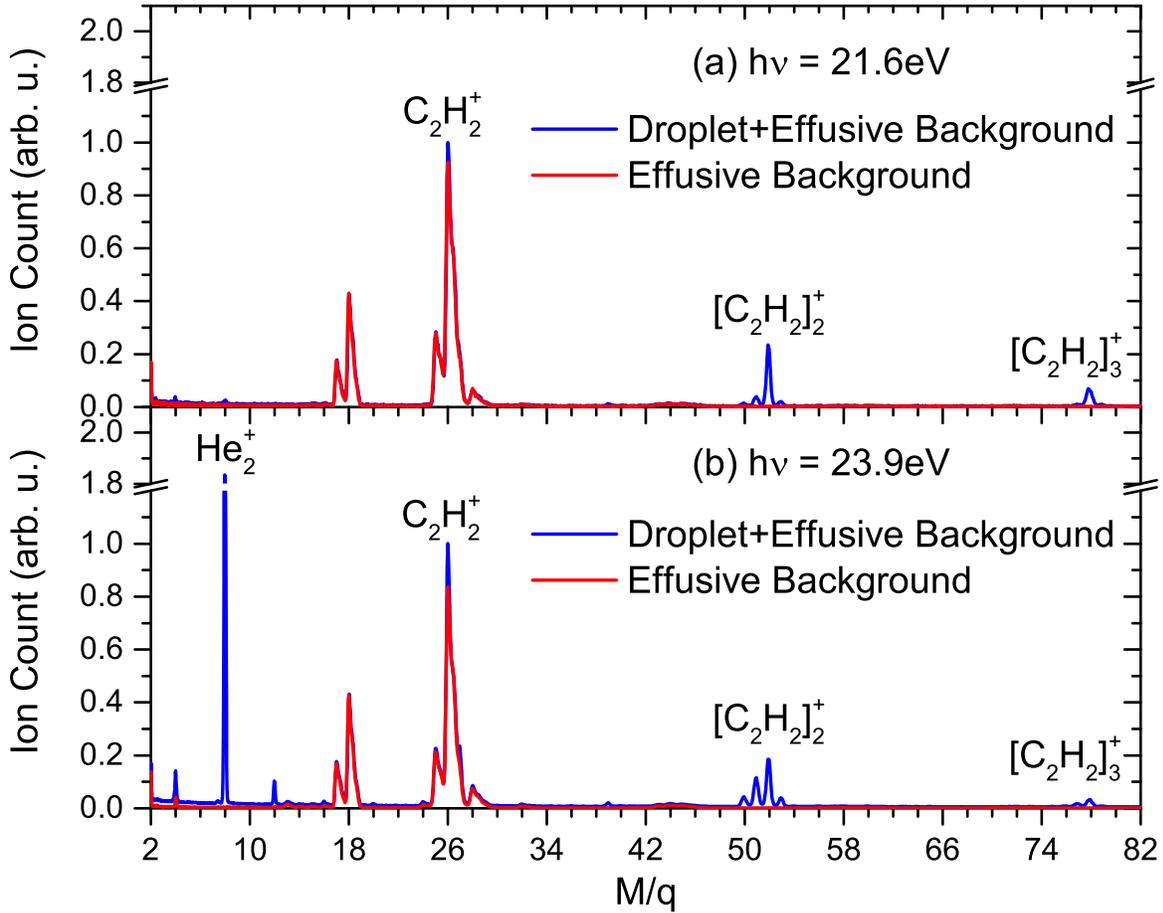}
	\caption{Ion mass spectra at photon energies of (a) $21.6$ \textrm{eV}, and (b) $23.9$ \textrm{eV} for $T_{noz}=14$ \textrm{K} and $P_{d}=4.5\times10^{-6}$ \textrm{mbar}. The blue line represents droplet and effusive background signal and red line represents only the effusive background signal. The horizontal axis shows the mass ($M$) to charge ($q$) ratio of the ionic fragments.}
	\label{Fig3_Ac}
\end{figure}

Fig.\ref{Fig3_Ac} shows the ion mass spectra at two different photon energies: a) $21.6$ \textrm{eV} and b) $23.9$ \textrm{eV}, where the red curve corresponds to the ionization of residual background gases and the blue curve is the signal from droplet ionization in addition to the background. The ion mass spectra are normalized such that the background \ce{N2^+} yields are proportional to the photoionization cross sections of \ce{N2} for producing \ce{N2^+} at the respective photon energies \cite{fen92}. We observe acetylene oligomer ions including the monomer (\ce{C2H2^+}), the dimer (\ce{[C2H2]_2^+}) and the trimer (\ce{[C2H2]_3^+}) from doped droplets. At  $23.9$ \textrm{eV}, we also observe \ce{He} cluster ions ($\ce{He}_{\mathit{m}}^{+}$, $m=1-3$) originating from droplet autoionization. We also observe more extensive fragmentation of acetylene oligomer ions  around the corresponding dimer and trimer ion peaks at this photon energy which is significantly different from the Penning ionization at $h\nu = 21.6 \mathrm{eV}$. The fragmentation we observe in the charge-transfer process occurring in the droplet is reminiscent of that reported in the case of a similar process in thermal collisions of \ce{He^+} ions with \ce{C2H2} molecule \cite{kim75}.
	
In the Penning ionization process, at $h\nu = 21.6$ \textrm{eV}, acetylene dimer and trimer ions along with their fragments were detected from droplet ionization. But the yield of monomer \ce{C2H2^+} ions with significant contrast of droplet specific signal over residual gas background was extremely low. Both for single doping ($P_{\textrm{d}}=6.5\times10^{-7}$ \textrm{mbar}) and for conditions optimized to multiply dope the \ce{He} droplet with \ce{C2H2} molecules with relatively high acetylene pressures ($P_{\textrm{d}}=4.5\times10^{-6}$ \textrm{mbar}) in the doping chamber, the signal of droplet specific monomer \ce{C2H2^+} ions was low. Note that, even for multiple doping conditions, only acetylene monomers are doped into the \ce{He} droplet. This is evident from the residual background ion signal where only \ce{C2H2} monomer signal is present. We ascribe two reasons for the low detection of \ce{C2H2} monomer ions compared to \ce{C2H2} oligomer ions in this Penning ionization process:
a) Suppression of the escape of the smaller ion (\ce{C2H2^+}) from the  droplet following the Penning ionization; b) In the case of multiply doped droplets, the formation of larger oligomer ions by the association of first Penning ionized \ce{C2H2^+} with other doped neutral \ce{C2H2} molecules in the droplet. We will discuss this further in the context of the PIES in the next section. The energy released in the association process evaporates several \ce{He} atoms from the droplet leading to a disintegration of the complex which enables the escape and eventual detection of \ce{C2H2} oligomer ions.
	
\subsection{Electron Energy Spectra}
\subsubsection{At $n=2$ droplet excitation band}

\begin{figure}[]
	\centering
	\includegraphics[width=0.6\textwidth]{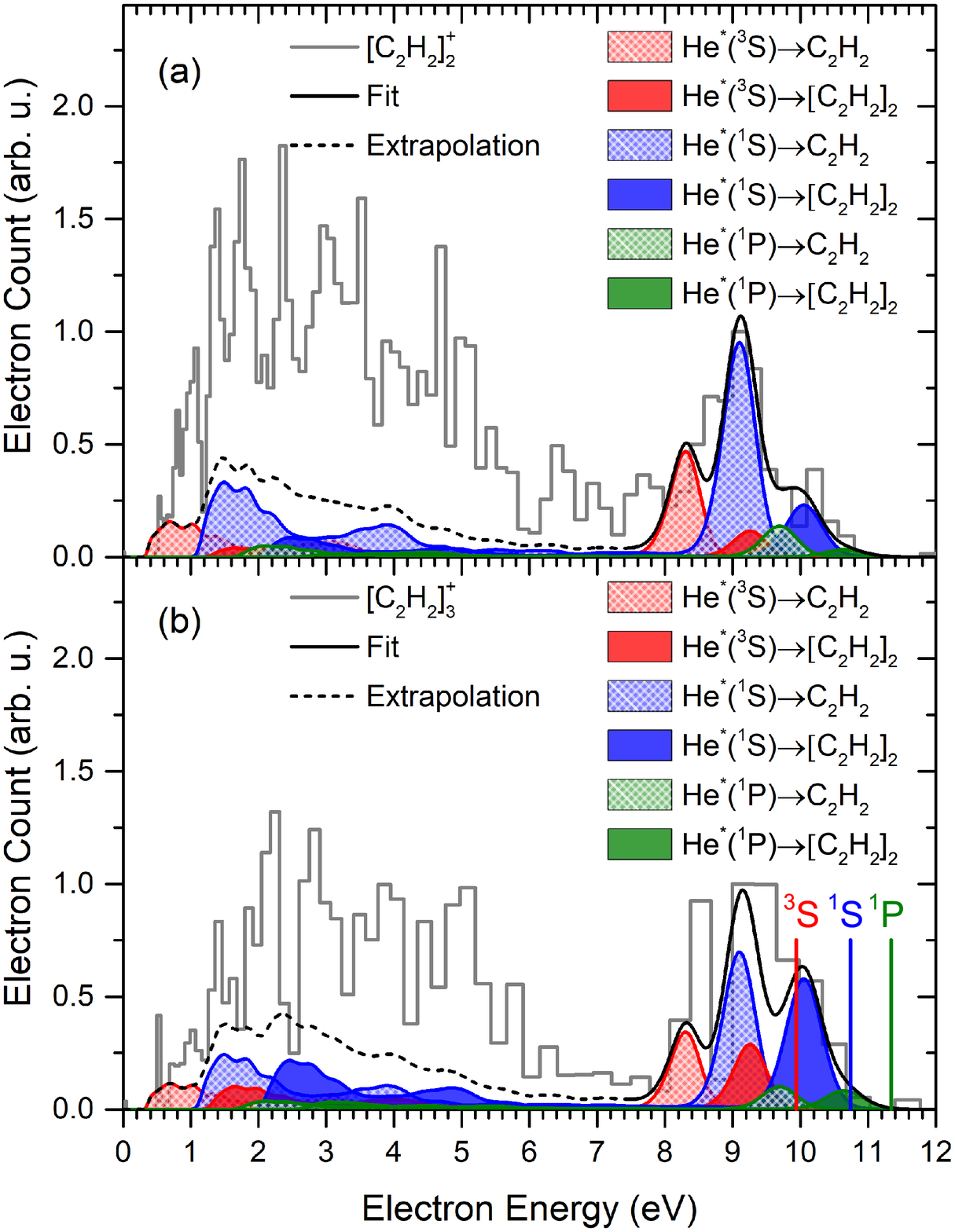}
	\caption{PIES correlated to (a) \ce{[C2H2]_2^+} and (b) \ce{[C2H2]_3^+} from acetylene doped \ce{He} droplet ionization at photon energy of $21.6$ \textrm{eV} for $P_{d}=4.5\times10^{-6}$ \textrm{mbar} and $T_{noz}=14$ \textrm{K}. The red, blue and green spectra with hatched shading are the convoluted PIES for \ce{C2H2} monomer from \ce{He}$^\ast$ $1s2s$ $^{3}S$, $1s2s$ $^{1}S$ and $1s2p$ $^{1}P$ states, respectively, while the convoluted PIES of \ce{C2H2} dimer from these \ce{He}$^\ast$ states are represented by the red, blue and green spectra with filled shading, respectively. The solid black line in each panel shows the total fit performed over the high energy feature from $7.5$ \textrm{eV} to $11$ \textrm{eV} which is arising from the cationic states of \ce{C2H2} monomer and dimer correlated to \ce{C2H2^+} $X^{2}\Pi_{u}$ state. The parameters for the fit were determined only using experimental data in the region $7.5$ \textrm{eV} and beyond. Nonetheless, we extrapolate this using the dashed black line into the region $\leq 7.5$ \textrm{eV} considering the corresponding $A^{2}\Sigma_{g}^{+}$ and $B^{2}\Sigma_{u}^{+}$ states. In panel (b), the red, blue and green vertical lines denote the PIES peak positions of \ce{C2H2} trimer Penning ionization from \ce{He}$^\ast$ $1s2s$ $^{3}S$, $1s2s$ $^{1}S$ and $1s2p$ $^{1}P$ states, respectively.} 
	\label{Fig4_Ac}
\end{figure}

Fig.\ref{Fig4_Ac} presents the PIES correlated to (a) acetylene dimer ions (\ce{[C2H2]_2^+}) and (b) acetylene trimer ions (\ce{[C2H2]_3^+}) originating from dopant Penning ionization when the doped droplet is photoexcited at $21.6$ \textrm{eV}. The measured PIES in coincidence with the dimer and trimer ions are quite similar and consist of two broad features - the first, between $0.5$ and $7.5$ \textrm{eV} and another in the range $7.5-11$ \textrm{eV}.
	 
Earlier reports on rare gas \cite{wan08} and alkali metal doped droplets \cite{buc13, lta19}, provided evidence that upon photoexcitation at $21.6$ \textrm{eV}, the dopants are Penning ionized not only from the \ce{He}$^\ast$ $1s2p$ $^{1}P$ dipole excited state but also prominently from long-lived  \ce{He}$^\ast$ $1s2s$ $^{3,1}S$ states which are populated upon fast relaxation. We  interpret the measured PIES correlated to acetylene dimer and trimer ions using PIES arising out of all these channels.
	
\citeauthor{ohn83} \cite{ohn83} studied Penning ionization of \ce{C2H2} by metastable \ce{He}$^\ast$ ($1s2s$ $^{3}S$) in slow atomic collisions and reported the corresponding PIES. To obtain the PIES of acetylene monomer due to Penning ionization by \ce{He}$^\ast$ in the $1s2s$ $^{1}S$ and the dipole allowed $1s2p$ $^{1}P$ states, we shifted the reported PIES by the corresponding energy differences, $+0.8$ \textrm{eV} and $+1.4$ \textrm{eV}, respectively. The relative amplitudes of the three features are freely adjusted in the fitting of high energy feature of the measured PIES, though. Noting that the ionization threshold of \ce{C2H2} dimer lies $0.96$ \textrm{eV} below the ionization threshold of its monomer \cite{ono82a}, the PIES of the \ce{C2H2} dimer is derived from the corresponding PIES of the monomer by a further shift of $+0.96$ \textrm{eV}. For the sake of keeping the number of free parameters within a reasonable limit, the relative amplitudes of the three components in the model of the dimer spectrum are kept fixed at the values obtained from the fit using the monomer fit function. Due to finite energy resolution of the VMI spectrometer the derived PIES are convoluted with the spectrometer instrument function. The resulting convoluted PIES for \ce{C2H2} monomer Penning ionization from \ce{He}$^\ast$ $1s2s$ $^{3}S$, $1s2s$ $^{1}S$ and $1s2p$ $^{1}P$ states are represented as red, blue and green lines with hatched shading, respectively. The convoluted PIES of \ce{C2H2} dimer Penning ionization from these three states of \ce{He}$^\ast$ are shown as red, blue and green lines with filled shading, respectively.

The high energy feature ($7.5-11$ \textrm{eV}) correlated to \ce{[C2H2]_2^+} (cf. fig.\ref{Fig4_Ac}a) fits quite nicely to the modeled sum of PIES of the acetylene monomer and dimer. The contribution of the $1s2s$ $^{3}S$ state of \ce{He}$^\ast$ to the experimental PIES is relatively large ($\sim 28 \%$) compared to previous findings \cite{lta19}. While the $1s2s$ $^{3}S$ state is not expected to be efficiently populated by droplet-induced relaxation, other factors may enhance the corresponding measured electron signal in the PIES. The Penning ionization cross section as well as the ejection mechanism out of the droplet leading to bare \ce{[C2H2]_2^+} Penning ions also determine the amplitude of the $1s2s$ $^{3}S$-state contribution.

The predominant contribution of this well-matched fit comes from the three channels corresponding to the Penning ionization of the acetylene monomer which is significantly higher than that from the ionization of the dimer. Likewise, applying this procedure to PIES correlated with  \ce{[C2H2]_3^+} (cf. fig.\ref{Fig4_Ac}b), using the same relative amplitudes of Penning ionization channels from \ce{He}$^\ast$ ($1s2s$ $^{3}S$, $1s2s$ $^{1}S$, and $1s2p$ $^{1}P$) states obtained from the model-fitting of \ce{[C2H2]_2^+} PIES, results in a  higher contribution of the monomer ionization than the dimer ionization to match with the observed spectrum. If Penning ionization occurred directly from these trimers, we would have expected the corresponding peak structures in the PIES beginning from $\sim 10$ \textrm{eV} rather than $\sim 8$ \textrm{eV} as seen in fig.\ref{Fig4_Ac}, as the ionization threshold of free acetylene trimers is known to be $9.83$ \textrm{eV} \cite{ono82b}. This would result in maximum kinetic energies of Penning electrons extending up to $\sim 12 \mathrm{eV}$, well beyond the observed limit of $\sim 11 \mathrm{eV}$. The red, blue and green vertical lines in fig.\ref{Fig4_Ac}b denote the PIES peak positions of \ce{C2H2} trimer Penning ionization from \ce{He}$^\ast$ $1s2s$ $^{3}S$, $1s2s$ $^{1}S$ and $1s2p$ $^{1}P$ states, respectively. Assuming again a fixed ratio of peak amplitudes, the contribution of these peaks to the fit of the overall spectrum would be negligible, though.

The ratios of dimer to monomer ionization channels for the PIES correlated to \ce{[C2H2]_3^+} and  \ce{[C2H2]_2^+} are $0.92$ and $0.27$, respectively. In both cases, the major contributions to Penning ionization are from the monomer ion rather than from larger oligomers. This indicates that acetylene molecules form loosely bound oligomers in \ce{He} nanodroplets, which largely retain the electronic structure of the acetylene monomer. This makes a strong case for processes alternative to the direct Penning ionization of acetylene dimers and trimers as the underlying mechanism for dopant ionization. This finding is in line with recent results from infrared spectroscopic studies of acetylene dimer in \ce{He} nanodroplets \cite{bri18}. The following picture emerges consistent with the salient features of the observed PIES correlated to acetylene dimer and trimer Penning ions: Following Penning ionization from an excited \ce{He}$^\ast$ in the droplet, the acetylene monomer and dimer ions associate with additional dopant molecules in the droplet to form larger acetylene oligomer ions. The energy of formation released into the droplet aids ejection out of the \ce{He} nanodroplets and the detection of the released bare dimer and trimer ions. Free jet studies indicate that the \ce{[C2H2]_2^+} and \ce{[C2H2]_3^+} are covalently bound with substantial binding energies $>2$ \textrm{eV} \cite{ste17}. We may also note that in singly doped droplets a monomer \ce{C2H2^+} ion does not have the benefit of further association and may not escape the droplet at all due to electrostrictive forces which bind it strongly to the \ce{He} host. Thus, the structure of an acetylene cluster formed in \ce{He} nanodroplets is that of a loosely bound van der Waals molecular aggregate rather than a covalently bonded system. This is reminiscent of a  foam-like structure consisting of discrete units of dopant monomers and dimers rather than entire $[\ce{C2H2}]_{n}$ clusters, for $n \geq 3$. Nonetheless, this van der Waals  aggregate collapses into a larger bound cluster $[\ce{C2H2}]_{n}^{+}$ ion upon Penning ionization. This is very similar to the photoinduced collapse observed in the case of \ce{Mg}-doped \ce{He} nanodroplets \cite{prz08}. Likewise, ionization induced recombination has been observed for \ce{Cr} atoms attached to \ce{He} nanodroplets \cite{kau15}. This is a central result of this work demonstrating the implementation of the Penning spectroscopy in droplets to elucidate the structure of dopant aggregates.
	
In the remainder of this article we discuss the low-energy feature ($0.5-7.5$ \textrm{eV}) in the PIES at $21.6$ \textrm{eV} photon energy (cf. fig.\ref{Fig4_Ac}) as well as indirect dopant ionization in the autoionization regime. Previous studies with rare-gas dopants, \ce{Xe} and \ce{Kr}, by \citeauthor{wan08} \cite{wan08}, measured PIES of the dopants arising from the \ce{He}$^\ast$ $1s2p$ $^{1}P$ and $1s2s$ $^{1}S$ states in the droplet. These electrons undergo inelastic scattering inside the droplet and lose energy which leads to a low-energy feature in the PIES. Recently, PIES of acene doped \ce{He} nanodroplet \cite{shc18} were found to be massively broadened presumably due to the scattering and many-body interaction of the emitted Penning electron with the surrounding \ce{He}. Similar to that work, we account for the scattering effects on the Penning electrons in the droplet by performing a Monte Carlo simulation. However, despite incorporating all the features of  electron-\ce{He} scattering in our simulations we found that even for large droplet radius of $30$ \textrm{nm} (compared to $6.5$ \textrm{nm} in our experiments for $T_{noz}=14$ \textrm{K} and \ce{He} expansion pressure of $50$ \textrm{bar}) the simulated PIES do not agree with the low-energy part of the photoelectron energy spectra. While the high energy feature ($7.5-11$ \textrm{eV}) is result of Penning ionization of \ce{C2H2} to \ce{C2H2^+} ($X^{2}\Pi_{u}$), this low-energy feature ($0.5-7.5$ \textrm{eV}) in fig.\ref{Fig4_Ac} arises from the Penning ionization of \ce{C2H2} leaving it behind in higher excited $A^{2}\Sigma_{g}^{+}$ and $B^{2}\Sigma_{u}^{+}$ states whose ionization energies are $16.7$ \textrm{eV} and $18.8$ \textrm{eV}, respectively \cite{wel99}. This large Penning ionization signal involving the $A$ and $B$ states as compared to that in earlier molecular beams experiments \cite{ohn83} may be related to the steep dependence of the corresponding effective cross-sections on the collision energy \cite{hor06}.

\subsubsection{At $n=4$ droplet excitation band}

\begin{figure*}[]
    \centering
    \includegraphics[width=0.65\textwidth]{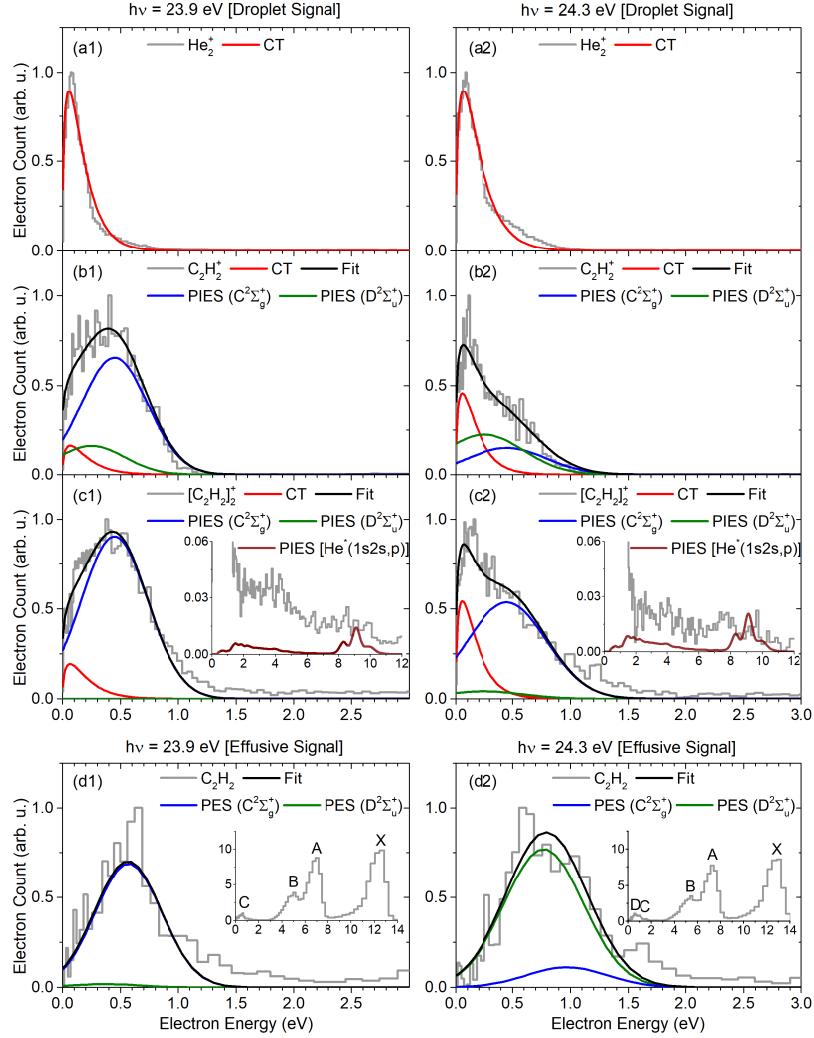}
    \caption{Electron energy spectra at two photon energies, (1) $23.9$ \textrm{eV} and (2) $24.3$ \textrm{eV}, correlated to different ions (a1,a2) \ce{He2^+}, (b1,b2) \ce{C2H2^+}, (c1,c2) \ce{[C2H2]_2^+} from the acetylene doped droplet ionization at $P_{d}=4.5\times10^{-6}$ \textrm{mbar} and $T_{noz}=14$ \textrm{K}. The conventional charge-transfer (CT) ionization processes are shown by red curves whereas the new Penning processes via \ce{He}$^\ast$ ($1s4p$ $^{1}P$) state leading to \ce{C2H2^+} $C^{2}\Sigma_{g}^{+}$ and $D^{2}\Sigma_{u}^{+}$ states are shown by blue and green curves respectively. Black curves are the sum of CT and Penning processes. The insets in panel(c1) and (c2) show the zoomed out electron spectra correlated to \ce{[C2H2]_2^+} ion at $h\nu=23.9$ \textrm{eV} and $h\nu=24.3$ \textrm{eV}, respectively, where we can see the presence of Penning ionization channels from \ce{He}$^\ast$ ($1s2s$, $1s2p$) states represented by brown lines in (c1,c2) observed at $h\nu=21.6$ \textrm{eV}. PES correlated to effusive \ce{C2H2} photoionization at (d1) $23.9$ \textrm{eV} and (d2) $24.3$ \textrm{eV} photon energies. The blue and green curves represent the simulated PES from \ce{C2H2^+} $C^{2}\Sigma_{g}^{+}$ and $D^{2}\Sigma_{u}^{+}$ states, respectively. The insets in panel(d1) and (d2), PES peaks from \ce{C2H2^+} $X^{2}\Pi_{u}$, $A^{2}\Sigma_{g}^{+}$, $B^{2}\Sigma_{u}^{+}$, $C^{2}\Sigma_{g}^{+}$ and $D^{2}\Sigma_{u}^{+}$ states can be observed.}
	\label{Fig5_Ac}
\end{figure*}

To learn about the ionization mechanisms in the autoionization regime ($23$ \textrm{eV} $<h\nu<E_{i}^{\ce{He}}$), we photoexcited the doped droplet across the $n=4$ droplet band corresponding to the atomic \ce{He}$^\ast$ ($1s4p$ $^{1}P$) excitation. The electron spectra correlated to \ce{He2^+}, \ce{C2H2^+} and \ce{[C2H2]_2^+} are presented in fig.\ref{Fig5_Ac} at two different photon energies, $23.9$ \textrm{eV} where a pronounced ionization maximum was observed, and $24.3$ \textrm{eV} which is a minimum that follows in fig.\ref{Fig2_Ac}.
	
	
The electron spectra correlated to \ce{He2^+} at both photon energies (cf. fig.\ref{Fig5_Ac}a1,a2) reveal very low-energy ($\sim 100 - 500$ \textrm{meV}) electrons which is a signature of \ce{He} nanodroplet autoionization \cite{buc13b}. Interestingly, the electron spectra correlated to \ce{C2H2^+} and \ce{[C2H2]_2^+} (cf. fig.\ref{Fig5_Ac}b1,b2,c1,c2) at $23.9$ \textrm{eV} and $24.3$ \textrm{eV} photon energies also show similar low energy features although these are broader with their maxima shifted to higher energies as compared to that of \ce{He2^+}. This indicates that conventional charge-transfer from the \ce{He2^+} in the droplet formed due to autoionization  \eqref{CT_ionization} cannot be the only mechanism for the underlying formation of the acetylene oligomer ions. The additional low energy peak around $0.5$ \textrm{eV} could therefore be due to Penning ionization of \ce{C2H2}, leading to  highly excited \ce{C2H2^+}. We measured photoelectron spectra (PES) from effusive \ce{C2H2} molecule at the two relevant photon energies (cf. fig.\ref{Fig5_Ac}d1,d2). Simulated PES from the higher-lying $C^{2}\Sigma_{g}^{+}$ and $D^{2}\Sigma_{u}^{+}$ states of \ce{C2H2^+} whose ionization energies are $23.33$ \textrm{eV} and $23.53$ \textrm{eV} \cite{wel99}, respectively, fit the observed PES quite well.


To confirm the additional Penning ionization of acetylene from $1s4p$ $^{1}P$ state of \ce{He}$^\ast$ resulting in \ce{C2H2^+} in $C^{2}\Sigma_{g}^{+}$ and $D^{2}\Sigma_{u}^{+}$ states, we used the following two-function fitting procedure to analyse the electron spectra correlated to \ce{C2H2^+} and \ce{[C2H2]_2^+} from droplet ionization:
$$F(E)=\frac{C^{MB}}{(k_{B}T)^{3/2}}\sqrt{E}e^{-\frac{E}{k_{B}T}}+\sum_{i=1,2}\frac{C^{G}_{i}}{\sigma}e^{-\frac{(E-E^{p}_{i})^{2}}{2\sigma^{2}}}$$
where the Maxwell-Boltzmann distribution function (first term) represents the charge-transfer ionization component and the Gaussian functions (second term) represent the new Penning ionization channels. Using the value of $T$, obtained from the fitting of the electron spectra in coincidence with \ce{He2^+}, we fit the electron spectra correlated to \ce{C2H2^+} and \ce{[C2H2]_2^+} varying the coefficients $C^{MB}$ and $C^{G}_{i}$. Since Penning ionization could occur from atomic \ce{He}$^\ast$ ($1s4p$ $^{1}P)$, the Gaussian peak positions $(E^{P}_{i})$ are fixed at the energy difference between \ce{He}$^\ast$ ($1s4p$ $^{1}P)$ and \ce{C2H2^+} ($C^{2}\Sigma_{g}^{+}$, $D^{2}\Sigma_{u}^{+}$) states while the standard deviation ($\sigma$) is fixed from the fitting of the effusive acetylene. As the energy distribution of the autoionized electrons which are detected in coincidence with \ce{He2^+} fits the Maxwell-Boltzmann distribution quite well, we used the same Maxwell-Boltzmann distribution to fit the charge-transfer ionization component. Note that, the observed energy distribution of the autoionized electrons are relatively high compared to previous report by \citeauthor{pet03} \cite{pet03} due to the finite energy resolution of the VMI.

The charge-transfer components are marked by red curves whereas blue and green lines represent the new Penning ionization components from \ce{C2H2^+} $C^{2}\Sigma_{g}^{+}$ and $D^{2}\Sigma_{u}^{+}$ states, respectively, in fig.\ref{Fig5_Ac} b1), b2), c1) and c2). Thus, we identify a prominent Penning mechanism leading to higher-lying $C^{2}\Sigma^{+}_{g}$ and $D^{2}\Sigma^{+}_{u}$ states of \ce{C2H2^+} via \ce{He}$^\ast$ $1s4p$ $^{1}P$.

Whereas at $h\nu=23.9$ \textrm{eV} the Penning channel dominates, at $h\nu=24.3$ \textrm{eV} charge-transfer ionization is more favourable. We notice that the electron spectra in coincidence with \ce{[C2H2]_2^+} at these energies (insets of fig.\ref{Fig5_Ac}c1, c2) also have weak long tail, extending up to nearly $11$ \textrm{eV}. This tail is possibly due to Penning ionization from \ce{He}$^\ast$ $1s2s$, $1s2p$ states arising from internal relaxation, as observed in PIES at $h\nu = 21.6$ \textrm{eV}.

Upon photoexcitation to the $n=2$ droplet excitation band at $21.6$ \textrm{eV}, we observed Penning ionization of acetylene clusters leading to \ce{C2H2^+} in $X$, $A$, and $B$ states. However, when the droplet is photoexcited to even higher $1s4p$ state, Penning ionization channels leading to higher-lying \ce{C2H2^+} states such as $C$ and $D$ states become energetically accessible and are observed here in this autoioization regime along with Penning ionizations leading to lower-lying \ce{C2H2^+} states ($X,A,B$). The enhanced Penning ionization cross-section leading to $C$ and $D$ states by \ce{He}$^\ast$ $1s4p$ $^{1}P$ could be related to the near degeneracy of these $C$ and $D$ states with the $1s4p$ $^{1}P$ state.

This interpretation is further supported by PIES recorded for \ce{He} nanodroplets doped with \ce{Li} atoms presented previously by \citeauthor{lta19} \cite{lta19}. The spectrum recorded at $h\nu=21.6$ \textrm{eV} (maximum of the $1s2p$ $^{1}P$ absorption band) is dominated by electrons emitted by Penning ionization of \ce{Li} interacting with excited \ce{He} in the $1s2s$ $^{1}S$ and $^{3}S$ states. In the coincident electron energy spectra recorded above the droplet ionization threshold ($23$ \textrm{eV}), the peaks between $13$ and $16$ \textrm{eV} reflects Penning ionization of \ce{Li} after electronic relaxation of the excited \ce{He} droplet into the $1s2s$ $^{1}S$ and $^{3}S$ states \cite{lta19}, and the low-intensity feature around $18$ \textrm{eV} is due to Penning ionization involving the $1s3s$, $1s3p$ and $1s4s$, $1s4p$ states of \ce{He}$^\ast$. The fact that the contribution to the \ce{Li} Penning ionization signal arising from these higher excited states is significantly lower than in the case of acetylene may be due to the differing location of the dopants in the droplet: \ce{Li} atoms are on the surface, whereas \ce{C2H2} molecules are expected to be located in the \ce{He} droplet interior. Consequently, a \ce{He}$^\ast$ excitation initially localized within the \ce{He} droplet migrates over a significant distance until it reaches the \ce{Li} dopant which affords the relaxation into the metastable $1s2s^{1}S$ and $^{3}S$ states before Penning ionization occurs. In comparison, the distance between the \ce{He}$^\ast$ and the acetylene dopant is shorter on average thereby facilitating direct Penning ionization prior to \ce{He}$^\ast$ relaxation.

\section{Conclusion}

Scattering of electrons following the Penning ionization of dopants in \ce{He} nanodroplets is thought to obscure molecular electron spectra, as demonstrated earlier in the case of acene molecules used as dopants \cite{shc18}. In this work we show that this is not always the case. Penning ionization can indeed be used as spectroscopic tool to study atomic and molecular quantum aggregates formed in \ce{He} nanodroplets by exciting the host matrix. By studying the electrons and ions for the host and dopant in coincidence, we identify relevant excited states of the host and the dopant. Furthermore, we demonstrate that this generic Penning ionization electron spectroscopy scheme is not limited only to the $n = 2$ droplet excitation but can be extended to perform spectroscopy employing higher droplet excitation bands such as $n = 4$ to probe the excited states of the dopant cluster e.g., the $C$ and $D$ states in acetylene ion. Employing this technique we uncover the structure of acetylene clusters formed inside nanodroplets. They coalesce in the form of loosely bound van der Waals aggregates, akin to a foam as observed for magnesium, rather than as a covalently bound system. This structure collapses into a composite oligomer ion following Penning ionization. Our work motivates further investigation of the structure of molecular aggregates in \ce{He} nanodroplets. This study establishes Penning ionization electron spectroscopy as a widely applicable technique to probe mesoscopic quantum aggregates in nanodroplets which behave like nano-cryostats even when they are photoactivated.

\section*{Conflicts of interest}
There are no conflicts to declare.

\section*{Acknowledgements}
VS, RG and SRK are grateful to DST, India and ICTP, Trieste, for support (proposal \# 20165468) to carry out this campaign at  the Elettra Synchrotron facility. VS and SRK acknowledge financial support from the IMPRINT scheme. SRK thanks the Max Planck Society for supporting this research via the Partner group. MM acknowledges support from the Carlsberg Foundation, and with SRK for the funding from the SPARC programme, MHRD, India.

\bibliography{Arxiv_Acetylene_Mandal} 
\bibliographystyle{unsrt}
\end{document}